# Fermilab 4.3-MeV Electron Cooler

Sergei Nagaitsev, Lionel Prost, and Alexander Shemyakin

Fermi National Accelerator Laboratory, Batavia, IL 60510

**Introduction**

The antiproton source for a proton-antiproton collider at Fermilab was proposed in 1976 [1]. The proposal argued that the requisite luminosity (~$10^{29}$ cm$^{-2}$s$^{-1}$) could be achieved with a facility that would produce and cool approximately $10^{11}$ antiprotons *per day*. At the end of its operation in 2011, the Fermilab antiproton production complex consisted of a sophisticated target system, three 8-GeV storage rings (namely the Debuncher, the Accumulator and the Recycler), 25 independent multi-GHz stochastic cooling systems and the world's only relativistic electron cooling system. Sustained accumulation of antiprotons was possible at the rate of greater than $2.5 \times 10^{11}$ *per hour*.

The production of antiprotons started with a 120-GeV proton beam from the Main Injector striking an Inconel target every 2-3 seconds. From all the particles thus created, 8.9–GeV/c antiprotons were collected in the Debuncher and stored in the Accumulator (a.k.a. stacking). The Accumulator antiproton stack was periodically transferred to the Recycler [2] where electron cooling allowed for a much larger antiproton intensity to be accumulated with smaller emittances. Typically 22-25$\times 10^{10}$ antiprotons were transferred to the Recycler every ~60 minutes. Prior to electron cooling in the Recycler, antiprotons destined for the Tevatron were extracted from the Accumulator only. Since late 2005, all Tevatron antiprotons were extracted from the Recycler only, which directly allowed for significant improvements in Tevatron luminosity. Figure 1 illustrates the flow of antiprotons between the Accumulator, Recycler and Tevatron over a one-week period.

In this paper we will briefly describe the Recycler ring, its electron cooling system, and the physics principles of electron cooling; then, we will present the two main types of measurements used to characterize and to tune the electron beam; finally, we will also discuss the optimization strategy for cooling in the context of maximizing the collider integrated luminosity.

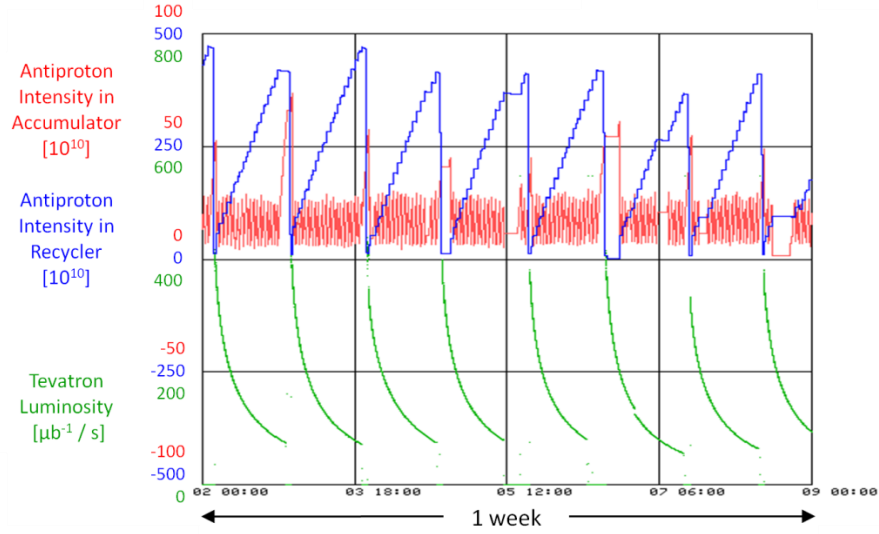

**Figure 1:** The production and the transfers of antiprotons between the Accumulator and the Recycler over a period of one week. While the Tevatron had a colliding beam store, small stacks of antiprotons were produced and stored in the Accumulator, and then periodically transferred to the Recycler in preparation for the subsequent Tevatron fill.

**Recycler ring**

The Recycler is a permanent-magnet, fixed-momentum (8.9 GeV/c) storage ring located in the Main Injector tunnel. Its main parameters are summarized in Table 1.

**Table 1:** Recycler ring main parameters

|  | Units | Value |
| --- | --- | --- |
| Circumference | m | 3320 |
| Acceptance (norm) | π mm mrad | 40 |
| Fractional momentum aperture | % | ±0.25 |
| Maximum dispersion function | m | 2 |
| Average $\beta_f$ | m | 40 |
| Average beam pipe radius | mm | 23 |
| Beam momentum | GeV/c | 8.9 |
| Average beam relativistic $\gamma$ |  | 9.48 |

The Recycler was designed to provide storage for a very large numbers of antiprotons (up to $6\times10^{12}$) and increase the effective antiprotons production rate by recapturing unused antiprotons at the end of collider stores (hence the name Recycler). Recycling of antiprotons was determined to be ineffective and was never implemented. However, the Recycler was used as a final antiproton cooling and storage ring. As such, the Recycler objectives were to accumulate and store antiprotons with high efficiency (primarily low beam loss during injection/extraction and high lifetime), allow for fast and frequent transfers of antiprotons from the Accumulator, and provide bunches with low emittance to the Tevatron. To meet these requirements, general machine improvements, elaborate beam and RF manipulations, and appropriate cooling schemes were developed and implemented [3].

The Recycler had a number of stochastic cooling systems in operation from day one; the electron cooling system had been envisioned as an upgrade [2] to complement the stochastic cooling system (in particular the longitudinal one because of the longitudinal injection scheme in the Recycler) and was placed into operation within days of its first successful demonstration in July 2005 [4]. With it, the Recycler has been able to store up to $6\times10^{12}$ antiprotons with acceptable lifetime (200-1000 hours). In routine operations, for which the preeminent figure of merit was the *integrated* luminosity rather than the number of antiprotons available for a store, the Recycler accumulated $3.5\text{-}4.0\times10^{12}$ antiprotons with a ~200-hr lifetime before injection into the Tevatron [5].

**Recycler Electron Cooling System**

Electron cooling is a method of increasing the phase-space density of "hot" heavy charged particles, ions or antiprotons, through Coulomb interactions with a "cold" electron beam, co-propagating with the same average speed in a small section of a ring. The method was proposed by G. Budker in 1967 [6], successfully tested in 1974 with low-energy protons [7], and later implemented at a dozen of storage rings (see, for example, a review [8]) at non-relativistic electron energies, $E_e < 300$ keV.

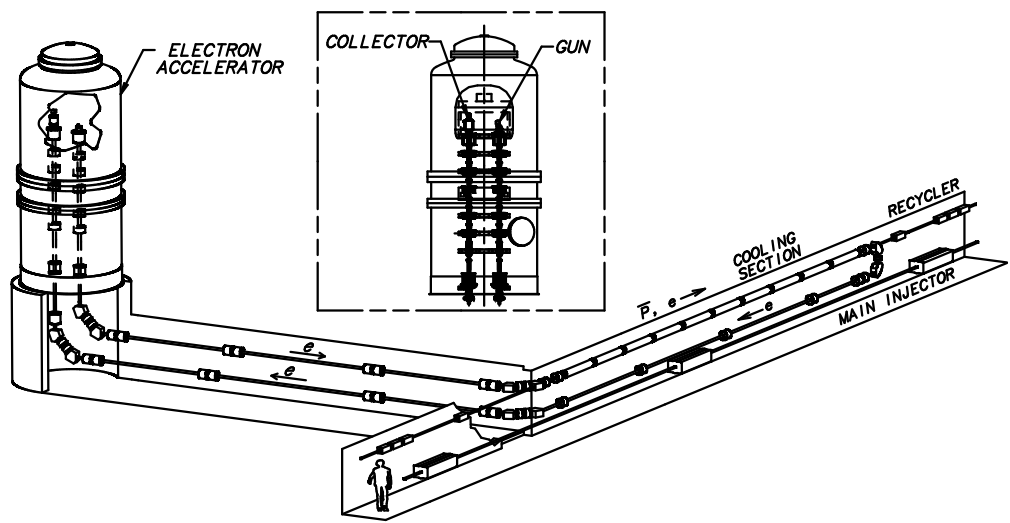

**Figure 2:** Schematic layout of the Recycler electron cooling system and the accelerator cross-section (inset).

Figure 2 shows the schematic layout of the Fermilab electron cooling system. The Pelletron (an electrostatic accelerator manufactured by the National Electrostatics Corp.) provided a 4.3-MeV (kinetic) electron beam (up to 500 mA, DC) which overlapped the 8-GeV antiprotons circulating in the Recycler in a 20-m long section and cooled the antiprotons both transversely and longitudinally. Table 2 shows the cooler design parameters.

**Table 2**: Electron cooler design parameters

| Parameter | Value | Unit |
|---|---|---|
| Electron energy | 4.34 | MeV |
| Beam current (for cooling) | 0.1 | A |
| Terminal voltage ripple, rms | 250 | V |
| Cathode radius | 2.5 | mm |
| Magnetic field at the cathode | ≤ 600 | G |
| Cooling section (CS) length | 20 | m |
| Solenoid field in CS | 105 | G |
| Beam radius in CS | 3.3 | mm |

The dc electron beam was generated by a thermionic gun, located in the high-voltage terminal of the electrostatic accelerator. This accelerator was incapable of sustaining dc beam currents to ground in excess of about 100 µA. Hence, to attain an electron dc current of 0.1-0.5 A, a recirculation scheme was employed, in which the electron beam that has interacted with the antiprotons is decelerated to 3.5 keV and accepted into the collector, located in the high-voltage terminal of the Pelletron. The typical relative beam current loss in the system was $2\times10^{-5}$ [9]. Note that for commissioning purposes, the electron gun was capable of operating in a pulse mode (typically, 2 µs at 1 Hz).

The Fermilab cooler employed a unique beam transport scheme [10]. The electron gun was immersed in a solenoidal magnetic field, which created a beam with a large angular momentum. After the beam was extracted from the magnetic field (at ~300 kV) and accelerated to 4.3 MeV, it was transported to the 20-m long cooling section solenoid using lumped focusing elements (as opposed to low-energy electron coolers where the beam remains immersed in a strong magnetic field at all times). The cooling section solenoid removed this angular momentum, and the beam was made round and parallel such that the beam radius, $a$, resulted in the same magnetic flux, $Ba^2$, as at the cathode. The magnetic field, $B$, in the cooling section was low, ~100 G, therefore the kinetics of the electron-antiproton scattering was weakly affected by the magnetic field. On the other hand, the field was strong enough to suppress the electron angle increase due to space charge and drift instability due to image charges. The cooling section included a 100 pairs of dipole correctors (horizontal/vertical) to correct for the solenoids magnetic field imperfections.

The beam line comprised 30 Beam Position Monitors (BPM) pairs (horizontal/vertical), 11 in the cooling section alone. Each BPM had one of its electrodes biased either positively or negatively to a few hundred volts (100-300 V typically) to either block (near the entrance of the acceleration and deceleration columns at the bottom of the Pelletron vessel) or collect positive ions created from beam- background gas interactions. Other diagnostics mainly consisted in a set of 10 movable scrapers in the cooling section, a DC Current Transformer (DCCT), 2 Optical

Transition Radiation (OTR) monitors and a YAG crystal, these last two being used in the pulse mode only.

**Electron Cooling Formulae**

An antiproton moving in a free electron gas with a velocity distribution $f_e(\vec{v}_e)$ experiences a friction force, which in a model of binary collisions can be written following Ref. [7]:

$$\vec{F}(\vec{v}_a) = 4\pi n_e r_e^2 mc^2 \eta \int f_e(\vec{v}_e) L_C \frac{(\vec{v}_e - \vec{v}_a)c^2}{|\vec{v}_e - \vec{v}_a|^3} d^3v_e \, , \tag{1}$$

where $n_e$ is the electron density in the beam rest frame, $m$ is the electron rest mass, $r_e$ is the classical electron radius, $\vec{v}_a$ is the antiproton velocity, and $\eta = L/C$ indicates the portion of the ring circumference, $C$, occupied by the cooling section of length $L$. The Coulomb logarithm, $L_C$, is defined as

$$L_C = \ln\left(\frac{\rho_{max}}{\rho_{min}}\right), \tag{2}$$

with the minimum and the maximum impact parameters, $\rho_{min}$ and $\rho_{max}$, in the Coulomb logarithm defined as

$$\rho_{min} = \frac{r_e c^2}{(\vec{v}_a - \vec{v}_e)^2}, \quad \rho_{max} = \min\left\{R_D, a, |\vec{v}_a - \vec{v}_e|\cdot\tau\right\}. \tag{3}$$

The maximum impact parameter is determined by the electron beam radius, $a$, (typically the case in the Fermilab cooler), the Debye shielding radius $R_D$, or the relative displacement of the particles during the traverse time through the cooling section $\tau = \frac{L_{cs}}{\gamma\beta c}$, where $\gamma$ and $\beta$ are the relativistic Lorentz factors of co-propagating particles in the lab frame, whichever is the smallest. In this paper, the electron velocity distribution is assumed to be Gaussian in each plane. Note that if the variations of the Coulomb logarithm in the integrand of Eq. (1) can be neglected, $L_C$ can be removed from the integrand and the instantaneous cooling rates of an antiproton beam with a Gaussian velocity distribution can be expressed with elementary functions [9].

**Cooling Force and Cooling Rate Measurements**

There are mostly two ways of assessing the cooling efficiency of the electron cooling system. One is to measure *directly* the cooling force, which in turn provides details of the electron beam properties. The second is to measure the speed at which the longitudinal and transverse emittances decrease (i.e. cooling rates) and determine the equilibrium emittance values. For operation, the cooling rates were the principal figure of merit for characterizing the status of electron cooling and used as a 'standard' check whenever the cooling performance was suspected to be deteriorating.

*Drag rate/cooling force*

The cooling properties of the electron beam were investigated primarily with 'drag rate' measurements obtained via a voltage-jump method similar to the one used in the early age of electron cooling [11]: a "pencil" coasting antiproton beam is cooled to an equilibrium; then, the electron energy is changed by a jump, and the rate of change for the mean value of the antiproton beam momentum distribution is recorded, while the antiprotons are dragged toward the new equilibrium. If the momentum spread remains small in comparison with the difference between the two equilibriums, this '*drag* rate' is equal to the longitudinal *cooling* force. Results of the drag force measurements as a function of the voltage jump amplitude (expressed in units of the antiproton momentum offset) are presented in Fig. 3. For these data, the electron and antiproton beams were concentric and collinear, which was defined as the electron beam being 'on-axis'.

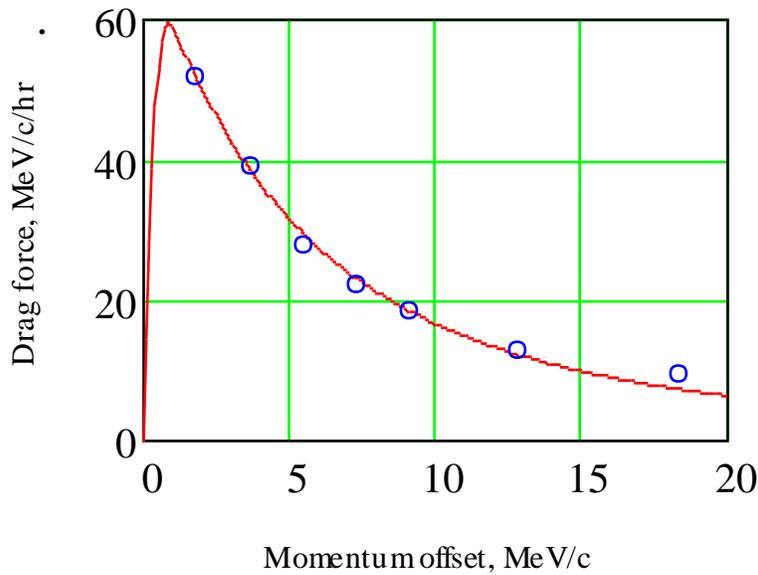

**Figure 3**: The drag rate on-axis as a function of antiproton momentum offset. The electron beam current was $I_e = 0.1$ A. The circles are data, and the solid line is a calculation using Eq. (1) with the rms electron divergence of 80μrad and the rms electron energy spread of 200 eV, $L_c = 9$.

For the case of the Fermilab cooler, the main contribution to the cooling force comes from collisions with low impact parameters. Therefore, the drag rate depends primarily on the electron beam properties in the vicinity of the probing antiproton beam. In turn, the information about the transverse distribution of the electron density and angles can be obtained with drag rate data taken at several spatial offsets (parallel to the beam axis) between the two beams in the cooling section. Fig. 4 shows an example of such measurements along with a fit to a simplified expression of the drag rate as a function of the transverse distance between the two beams (or equivalently, the radius of the electron beam) written as

$$F_{lz}(x) = F_0(\Delta p_a) \cdot \begin{cases} \dfrac{1-\left(\dfrac{x}{a}\right)^2}{1+\left(\dfrac{x}{b}\right)^2}, & x \leq a \\ 0, & x > a \end{cases}, \quad (4)$$

where $F_0$ is the maximum drag rate (by definition at the center of the electron beam current density transverse distribution) for a given momentum offset. In the fraction, the numerator approximates the electron beam density profile, determined from the electron gun simulations, while in the denominator, the parameter $b$ describes an increase of the electron angles with the radial offset. For such a profile, the finite size of the probe antiproton beam results in a decrease of the measured drag rate as compared to the cooling force experienced by the antiprotons on axis. The red curve in Figure 4 shows the corresponding correction.

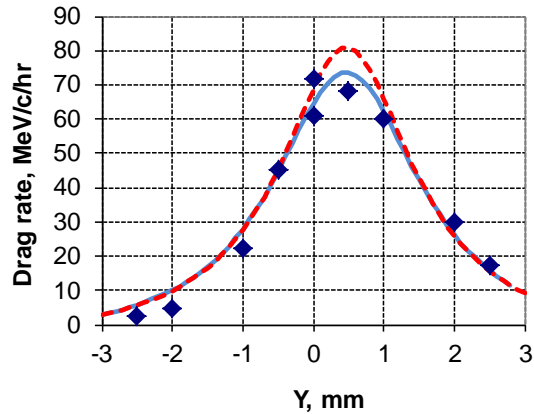

**Figure 4**: The drag rate as a function of the electron beam offset with respect to the co-propagating antiproton beam axis. The Pelletron voltage jump was 2 kV, $I_e = 0.3$ A, the number of antiprotons, $N_p = 1.3 \cdot 10^{10}$. The blue curve is the best fit to the model described with $a = 4.3$ mm, and the fitting parameters, $F_0 = 80$ MeV/c/hr and $b = 1.2$ mm. During the measurement, the rms size of the antiproton beam was estimated to be ~0.25 mm. The red dashed curve shows the fitted cooling force after correcting for the finite size of the antiproton beam.

If the electron angular spread remains constant, the cooling force should increase proportionally to the current density. Drag rates measured at different beam currents during the entire span of the cooler's operation are shown in Fig. 5 together with the simulated current density at the beam center.

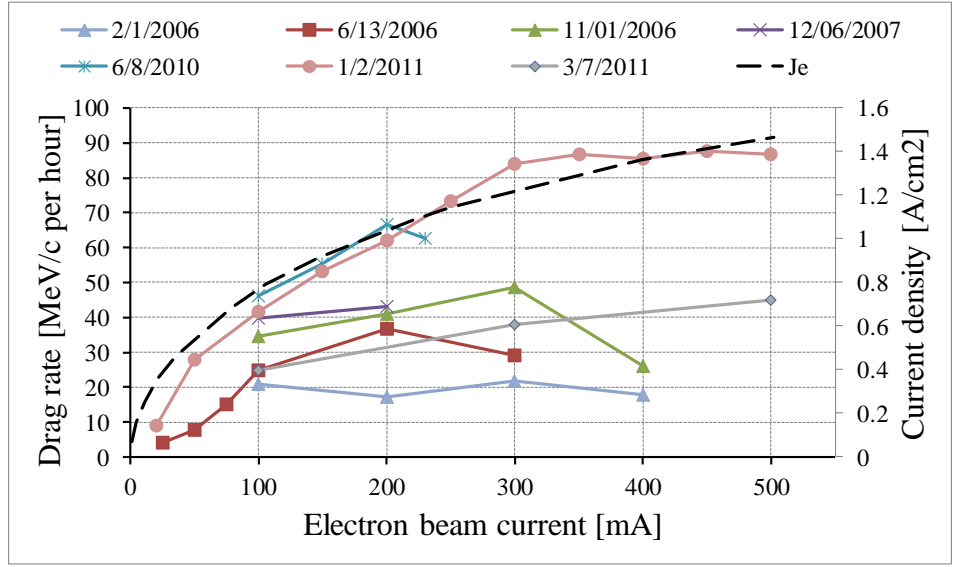

**Figure 5**: The drag rate measured on axis as a function of the beam current at various dates with a 2 kV voltage jump. The current density calculated at the beam center (dashed curve) is shown for comparison.

The large scatter in the measured drag rates is related to important variations of the electron angles in the cooling section. Until the end of the collider operation, significant efforts were devoted to understanding what determined these angles and how they could be reduced. Best estimates of the various contributions to the total rms electron angle are presented in Table 3.

**Table 3:** Contributions to the total rms electron angular divergence in the cooling section. Shown values are 1D, rms, obtained from averaging the angles over the cross section of a 0.1-A electron beam in the best scenario.

| Effect | Angle, μrad | Method of evaluation |
|---|---|---|
| Thermal velocities | 57 | Calculated from the cathode temperature |
| Envelope mismatch | ~50 | Resolution of tuning + optics simulations |
| Dipole motion (above 0.1 Hz) | ~35 | Spectra of BPMs in the cooling section |
| Dipole motion caused by field imperfections | ~50 | Simulation of electron trajectory in the measured magnetic field |
| Non-linearity of lenses | ~20 | Trajectory response measurements |
| Ion background | < 10 | Cooling measurements |
| **Total** | ~100 | Summed in quadratures |

With a detailed description of improvements and measurements given in Ref. [9], here we would like only to highlight several important milestones in the evolution of the electron beam angles:
- Quadrupole correctors allowed to significantly decrease the beam envelope angles at low beam currents.

- Development of a beam-based procedure for aligning the magnetic field in the cooling section alleviated the effect of mechanical drifts of the cooling section's solenoids.
- Clearing the background ions to <1% of the electron density by interrupting the electron beam for 2 µs at 100 Hz improved cooling at higher beam currents.

*Cooling rate*

While the drag rate measurements were the instrument to estimate and improve the electron beam properties, cooling efficiency for operation was described by the cooling rates. To measure cooling rates, the antiproton beam, confined by rectangular RF barriers, was first let diffuse for 15 minutes with no cooling (including stochastic cooling) and then the electron beam was turned on and cooled the antiprotons for 15 minutes. The cooling rate was calculated as the difference between the time derivatives of the momentum spread (or transverse emittances) before and after turning on the electron beam.

Typically, in this case the rms antiproton beam radius exceeded the size of the electron beam area with good cooling properties, and a model of cooling in an infinite homogenous electron gas predicted much higher cooling rates than were actually measured. One still can examine consistency between drag rates and cooling rates in a simple model assuming that measurements of the drag rates at various electron beam offsets (e.g. as in Fig. 4) represent the cooling force experienced by an antiproton at that given radius. Results of such comparisons are shown in Fig. 6, where cooling rates measured with similar electron beam conditions are plotted for different initial antiproton beam transverse emittances. The dash-dotted curve is the result of the integration of the cooling force, reconstructed from drag rate measurements for the same electron beam parameters at various offsets over a Gaussian spatial distribution of antiprotons with the rms size calculated from their measured emittance. Note that integration does not involve any additional fitting parameters. Taking into account the approximate nature of this model, the agreement is reasonable.

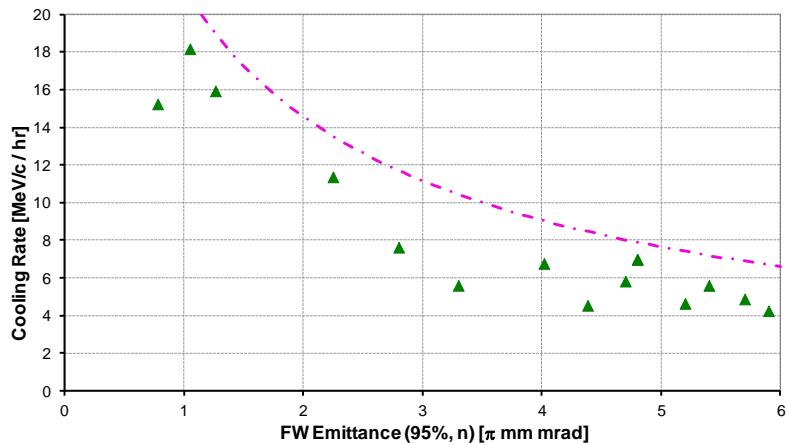

**Figure 6:** The longitudinal cooling rate as a function of the antiproton emittance for $I_e = 0.1$ A.

**Conclusion**

The Recycler Electron Cooler at Fermilab made an important contribution to the success of the Tevatron Run II by increasing the antiproton flux and brightness. It also marked a significant step in the development of accelerator technology and accelerator physics, demonstrating for the first time relativistic cooling as well as beam transport of a magnetized beam with lumped focusing.

Drag rate measurements proved to be the main tool for analyzing and improving cooling properties of the electron beam. Various types of cooling measurements were eventually found to be mutually consistent and in a reasonable agreement with a non-magnetized description of electron cooling.

## Acknowledgement

Fermilab is operated by Fermi Research Alliance, LLC under Contract No. DE-AC02-07CH11359 with the United States Department of Energy.

## References

[1] D. Cline et al., "Proposal to construct an antiproton source for the Fermilab accelerators", proposal 492, in Proceedings of 1976 NAL Summer Study on Utilization of the Energy Doubler/Saver, Fermilab, Batavia U.S.A. (1976), pg. 309.
[2] Fermilab Recycler Ring Technical Design Report, Ed. G. Jackson, Fermilab Preprint TM-1991 (1997).
[3] V.Lebedev, V.Shiltsev (Eds.), *Accelerator Physics at the Tevatron Collider* (Springer, New York, 2014)
[4] S. Nagaitsev et al., "Experimental Demonstration of Relativistic Electron Cooling", Phys. Rev. Lett. 96, 044801 (2006).
[5] A. Shemyakin and L.R. Prost, "Ultimate performance of relativistic electron cooling at Fermilab", in Proceedings of COOL11, THIOA01, Alushta, Ukraine (2011).
[6] G. Budker, Sov. Atomic Energy, vol. 22, p. 346, 1967.
[7] G. I. Budker et al., IEEE Trans. Nucl. Sci., Vols. NS-22, p. 2093, 1975.
[8] I. N. Meshkov, Phys. Part. Nucl., vol. 25, no. 6, p. 631, 1994.
[9] A. Shemyakin and L. Prost, "The Recycler Electron Cooler", FERMILAB-FN-0956-AD (2013), arXiv:1306.3175.
[10] A. Burov et al., "Optical principles of beam transport for relativistic electron cooling", Phys. Rev. ST, Accel. Beams 3, 094002 (2000).
[11] G.I. Budker et al., Preprint IYaF 76-32 (1976), http://cdsweb.cern.ch/record/118046/files/CM-P00100706.pdf